\documentclass[12pt]{iopart}
\begin{document}
\jl{6}
\newcommand{\vecp}[1]{\stackrel{\longrightarrow}{#1}}

\title[H.H. Fliche et al.]{Anisotropic Hubble expansion of large scale structures}

\author{H.H. Fliche\dag, J.M. Souriau\ddag\ and 
R. Triay\ddag\footnote[3]{triay@cpt.univ-mrs.fr}}

\address{\dag\ LMMT (UPRES~EA 2596) Fac. des Sciences et Techniques 
de St J\'er\^ome, av. Normandie-Niemen, 13397 Marseille Cedex 20, France}

\address{\ddag\  Centre de Physique Th\'eorique\footnote{Unit\'e Mixte de Recherche (UMR 6207) du CNRS, et des universit\'es Aix-Marseille I,
Aix-Marseille II et du Sud Toulon-Var. Laboratoire affili\'e \`a la FRUMAM (FR 2291).}, CNRS Luminy Case 907, 13288 Marseille Cedex 9, France }

\begin{abstract}
With the aim of understanding the cosmic velocity fields at large scale, we investigate the dynamics of a pressureless distribution of gravitational
sources moving under an anisotropic generalization of Hubble expansion and constraint by Euler-Poisson equations system. As a result, it turns out that
such a behavior requires the distribution to be homogenous, similarly to Hubble law.  Among several solutions, we show a planar kinematics with constant
(eternal) and rotational distortion, where the velocity field is not potential. Within this class, the one with no rotational distortion identifies to a
bulk flow. To apply this model within cosmic structures as the Local Super Cluster, the solutions are interpreted as approximations providing us with an
hint on the behavior of the cosmic flow just after decoupling era up to present date. Such a result suggests that the observed bulk flow may not be due
exclusively to tidal forces but has a primordial origin.
\end{abstract}


\submitted

\maketitle

\section{Introduction}
\label{Introduction}
In the past, the investigations of cosmic velocity fields from redshifts surveys were not as successful as that for providing us with powerful
tools for the estimation of cosmological parameters and the local density fields, see {\it e.g.\/} \cite{Strauss99}. Later on,  least action methods
\cite{Peebles89} were also used in reconstruction procedures with more reliable estimates, see {\it e.g.\/}
\cite{SusperregiBinney94,SchmoldtSaha98,Susperregi00,Goldberg00,GoldbergSpergel00}. Both approaches assume that the peculiar velocity field decomposes
into a ``divergent'' component due to density fluctuations inside the surveyed volume, and a tidal (shear) component, consisting of a bulk velocity and
higher moments, due to the matter distribution outside the surveyed volume \cite{CourteauDekel01}. These hypotheses are justified by the properties that
the irrotational linear perturbations dominate with time \cite{LifshitzKhalatnikov63, Fliche81} together with Kelvin theorem which ensures the irrotational
characteristics of motions. However, it turns out that theses investigations might provide us with biased results because of their inherent dependence of
sampling characteristics. For example, the presence of an high region (Great Attractor) was proposed for accounting of an unexpected feature in the
cosmic velocities \cite{LyndenBellEtal88,BertschingerDekel89,DekelEtal90,DekelEtal99}, when from recent surveys Shapley Concentration seems to be a better
candidate, although it is not responsible for all of the SMAC flow \cite{BranchiniEtal99,SaundersEtal00,HudsonEtal04}. Since such an interpretation is not
as clear cut as that, one might ask whether the presence of a bulk flow originates necessarily from a density excess in the spatial distribution of
gravitational structures or another alternative could be envisaged. For answering this question, we investigate Euler Poisson equations system
describing an anisotropic Hubble flow of pressureless distribution of gravitational sources in a Newtonian schema.

\section{Dynamics of the cosmic expansion}
\label{Dynamics}

The dynamics of a pressureless distribution of gravitational sources (dust) is investigated by assuming that
the motion of sources satisfy the following kinematics
\begin{equation}\label{HubbleLaw}
\vec{r}={\bf A}\vec{r}_{\circ},\qquad {\bf A}(t_{\circ})={\rm 1\mskip-4mu l}
\end{equation}
where ${\bf A}={\bf A}(t)$ stands for a $3\!\times\!3$ matrix depending on cosmic time $t$ which has to be determined by an observer at rest with respect
to Cosmological Background Radiation (CMB). The present investigation limits on {\em collisionless motions\/}, which is ensured by a non vanishing
determinant of matrix ${\bf A}$, and because it reads as a unit matrix at given $t=t_{\circ}$ one has ${\rm det}{\bf A}>0$. Hence, the velocity field
$\vec{v}=\vec{v}(\vec{r},t)$ is given by
\begin{equation}\label{velocity}
\vec{v}=\frac{{\rm d}\vec{r}}{{\rm d}t}=\dot{{\bf A}}{\bf A}^{-1}\vec{r},\qquad 
\dot{{\bf A}}=\frac{{\rm d}}{{\rm d}t}{\bf A}
\end{equation}
where ${\bf A}^{-1}$ stands for the inverse matrix (${\bf A}^{-1}{\bf A}={\bf A}{\bf A}^{-1}={\rm 1\mskip-4mu l}$). Moreover, accordingly to Hubble law,
we assume a {\em radial\/} acceleration field
\begin{equation}\label{acceleration}
\vec{g}=\frac{{\rm d}\vec{v}}{{\rm d}t}=\ddot{{\bf A}}{\bf A}^{-1}\vec{r},\qquad
\vec{g}\propto\vec{r}
\end{equation}
We assume now that these motions are constraint by Euler equations system
\begin{eqnarray}
\frac{\partial\rho}{\partial\,t}&=&-{\rm div}\left(\rho \vec{v}\right)\label{Euler2}\\
\frac{\partial \vec{v}}{\partial\,t}&=&-\frac{\partial\vec{v}}{\partial\vec{r}}
\vec{v}+\vec{g}\label{Euler1}
\end{eqnarray}
where $\rho=\rho(\vec{r},t)$ stands for the density. By using the trace, the determinant and the equality $\frac{\rm
d}{{\rm d}t} {\rm det}{\bf A}={\rm Tr}\left(\dot{{\bf A}}{\bf A}^{-1}\right){\rm det}{\bf A}$, eq.\,(\ref{Euler2}) transforms
\begin{eqnarray}
\dot{\rho}&=&\frac{\partial\rho}{\partial\,t}+\vec{v}\cdot\vecp{\rm grad}\rho \nonumber\\
&=&-\rho\;{\rm div}\left(\dot{{\bf A}}{\bf A}^{-1}\vec{r}\right)
=-\rho\;{\rm tr}\left(\dot{{\bf A}}{\bf A}^{-1}\right)
=-3\rho\;\frac{\dot{a}}{a}\label{gEuler2bis}
\end{eqnarray}
where the doted variables stand for time derivatives, and
\begin{equation}\label{a}
a(t)=\sqrt[3]{{\rm det}{\bf A}},\qquad a_{\circ}=a(t_{\circ})=1
\end{equation}
for the (generalized) {\em expansion factor\/}. Hence, with eq.\,(\ref{HubbleLaw}) one obtains
\begin{equation}\label{rhoEuler2}
\rho(\vec{r},t)=\frac{\rho({\bf A}^{-1}\vec{r},t_{\circ})}{a^{3}}=\frac{\rho_{\circ}}{a^{3}},
\qquad \rho_{\circ}=\rho_{\circ}(\vec{r}_{\circ})
\end{equation}
which simply accounts for the mass conservation $\rho_{\circ}{\rm d}r_{\circ}^{3}=\rho\,{\rm d}r^{3}=\rho\,{\rm det}{\bf A}\,{\rm
d}r_{\circ}^{3}$.

One has a unique matrix decomposition
 \begin{equation}\label{H}
\dot{{\bf A}}{\bf A}^{-1}=H{\rm 1\mskip-4mu l}+\frac{H_{\circ}}{a^{2}}{\bf B},\qquad H=\frac{\dot{a}}{a},\qquad H_{\circ}=H(t_{\circ}),\qquad {\rm
tr}\;{\bf B}=0
\end{equation}
where $H=H(t)$ acts as the usual Hubble factor and ${\bf B}={\bf B}(t)$ stands for a traceless matrix herein called {\em distortion matrix\/}. It
characterizes a deviation from isotropy of the (dimensionless) velocity field, its amplitude is defined by the matrix norm
 \begin{equation}\label{amplitude}
\|{\bf B}\|=\sqrt{{\rm tr}\left({\bf B}^{t}{\bf B}\right)}
\end{equation}
where the sign ''$^{\rm t}$'' stands for the matrix transposition. For convenience, let us write
 \begin{equation}\label{TrBn}
\beta_{n}(t)={\rm tr}\left({\bf B}^{n}\right),\qquad n=1,2,3
\end{equation}
According to eq.\,(\ref{velocity},\ref{Euler1},\ref{H}), one has
\begin{equation}\label{gEuler2}
\vec{g}
=\left(
\left(\dot{H}+H^{2}\right){\rm 1\mskip-4mu l}
+\frac{H_{\circ}}{a^{2}}\left(
\dot{\bf B}+\frac{H_{\circ}}{a^{2}}{\bf B}^{2}
\right)
\right)\vec{r}
\end{equation}
We assume that the gravitational field satisfies Poisson-Newton equations
\begin{eqnarray}
{\rm div}\vec{g} &=&-4\pi{\rm G}\rho+\Lambda \label{Poisson2}\\
\vecp{\rm rot}\vec{g} &=&\vec{0}\label{Poisson1}
\end{eqnarray}
where ${\rm G}$ is Newton constant of gravitation and $\Lambda$ the cosmological constant. According to eq.\,(\ref{gEuler2}), the left hand term of
eq.\,(\ref{Poisson2}) reads
\begin{equation}\label{TraceH}
{\rm div}\vec{g}={\rm tr}\left(\frac{\partial\vec{g}}{\partial\vec{r}}\right)=
3\left(\dot{H}+H^{2}\right)
+\frac{H_{\circ}^{2}}{a^{4}}\beta_{2}
\end{equation}
which does not depend on spatial coordinates. Hence, eq.\,(\ref{Poisson2}) tells us that the space distribution of sources is {\em homogenous\/}, {\em
i.e.\/}
\begin{equation}\label{homogene}
\rho=\rho(t)
\end{equation}
With eq.\,(\ref{Poisson2},\ref{TraceH}) one has
\begin{equation}\label{Poisson2b}
\frac{1}{H_{\circ}^{2}}\left(\dot{H}+H^{2}\right)
=\frac{1}{H_{\circ}^{2}}\frac{\ddot{a}}{a}
=-\frac{\Omega_{\circ}}{2a^{3}}+\lambda_{\circ}-\frac{1}{3a^{4}}\beta_{2}
\end{equation}
where
\begin{equation}\label{Parameters}
\Omega_{\circ}=\frac{8\pi G}{3H_{\circ}^{2}}\rho_{\circ},\qquad
\lambda_{\circ}=\frac{\Lambda}{3H_{\circ}^{2}},\qquad
H_{\circ}=H(t_{\circ})
\end{equation}
are motion parameters. By multiplying each term of
eq.\,(\ref{Poisson2b}) by
$2\dot{a}a$ one easily identifies the following constant of motion
\begin{equation}\label{k}
\kappa_{\circ}=\frac{\Omega_{\circ}}{a}+\lambda_{\circ}a^{2}-\frac{\dot{a}^{2}}{H_{\circ}^{2}}-\frac{2}{3}\int^{a}_{1}\frac{\beta_{2}}{a^{3}}{\rm d}a = \Omega_{\circ}+\lambda_{\circ}-1
\end{equation}
Hence, the chronology is given by
\begin{equation}\label{dt}
{\rm d}t=\frac{1}{H_{\circ}}\frac{a{\rm d}a}{\sqrt{P(a)}}
\end{equation}
where $H_{\circ}>0$ accounts for an expansion (according to observations, the case $H_{\circ}<0$ which accounts for a collapse is not envisaged), and
\begin{equation}\label{Pr}
P(a)=\lambda_{\circ}a^{4}-\kappa_{\circ}a^{2}+\Omega_{\circ}a-\frac{2}{3}a^{2}\int^{a}_{1}\frac{\beta_{2}}{a^{3}}{\rm d}a\geq
0,\qquad P(1)=1
\end{equation}
The constraint given by eq.\,(\ref{Poisson1}) can be written in matrix form as follows
\begin{equation}\label{Poisson1bis}
\frac{\partial\vec{g}}{\partial\vec{r}}- \left(\frac{\partial\vec{g}}{\partial\vec{r}}\right)^{\rm t}={\bf 0}
\end{equation}
Such a symmetric property with eq.\,(\ref{gEuler2},\ref{Poisson2b}) shows that the matrix 
\begin{equation}\label{Bstar}
\hat{{\bf B}}={\bf B}^{2}-\frac{1}{3}\beta_{2}{\rm 1\mskip-4mu l}+\frac{a^{2}}{H_{\circ}}\dot{\bf B}
\end{equation}
is traceless and symmetric
\begin{equation}\label{Bstar2}
{\rm tr}\hat{{\bf B}}=0,\qquad \hat{{\bf B}}^{\rm t}=\hat{{\bf B}}
\end{equation}
According to eq.\,(\ref{acceleration},\ref{gEuler2},\ref{Bstar},\ref{Bstar2}), since the field $\vec{g}$, which reads
\begin{equation}\label{g}
\frac{\vec{g}}{H_{\circ}^{2}}=\left(\lambda_{\circ}- \frac{\Omega_{\circ}}{2a^{3}}+\frac{1}{a^{4}}
\hat{{\bf B}}\right)\frac{\vec{r}}{a^{2}}
\end{equation}
is radial, the matrix $\hat{{\bf B}}$ must be scalar and because it is traceless, one has necessarily
\begin{equation}\label{Bstar0}
\hat{{\bf B}}={\bf 0}
\end{equation}

\subsection{Reference map}
Instead of $(t,\vec{r})$, it is more convenient to analyze the dynamics of the cosmic flow in the $(\tau,\vec{q})$ coordinates defined by
 \begin{eqnarray}
{\rm d}\tau&=&H_{\circ}\frac{{\rm d}t}{a^{2}}\label{u}\\
\vec{q}&=&\frac{\vec{r}}{a}\label{q}
\end{eqnarray}
herein called {\em reference map\/}. According to eq.\,(\ref{velocity},\ref{dt},\ref{Bstar},\ref{Bstar0}), the equations of motion read
 \begin{eqnarray}
\frac{{\rm d}\vec{q}}{{\rm d}\tau}&=&{\bf B}\vec{q},\qquad
\frac{{\rm d}^{2}\vec{q}}{{\rm d}\tau^{2}}=\frac{a^{2}}{3H_{\circ}}\beta_{2}\vec{q}\label{qvelocity}\\
{\rm d}\tau&=&\frac{{\rm d}a}{a\sqrt{P(a)}}\label{du}\label{t}
\end{eqnarray}
where the distortion matrix ${\bf B}$ satisfies
\begin{equation}\label{Bstar00}
\frac{{\rm d}{\bf B}}{{\rm d}\tau}= \frac{1}{3}\beta_{2}{\rm 1\mskip-4mu l}-{\bf B}^{2}
\end{equation}
The resolution of these equations can be performed by mean of numerical techniques; having solved
eq.\,(\ref{Bstar00}), which gives the evolution with time of distortion matrix ${\bf B}$, the particles trajectories $\tau\mapsto\vec{q}(\tau)$ are
obtained by integrating eq.\,(\ref{qvelocity}) and the evolution of the generalized expansion factor $a$ from eq.\,(\ref{t}).

\subsection{Discussion}
As a result, it is interesting to mention that (as it is the case for Hubble law) this anisotropic generalization accounts for homogeneous space
distributions of matter. Hence, one is forced to ask whether it describes correctly the dynamics of cosmic structures because of the presence of strong
density inhomogeneities in the space distribution of galaxies catalogs. In principle, such a remark should be also sensible to question Hubble
law when, regardless the isotropy, it is a fact that perturbations are not so dominant otherwise it would never have been highlighted. Actually,
homogeneity is implicitly assumed for the interpretation of CMB isotropy and the redshift of distant sources, which provides us with an expanding
background. Namely the comoving space of FL world model onto which the gravitational instability theory is applied for understanding the formation of
cosmic structures. It is with such a schema in mind that this anisotropic Hubble law may provides us with an hint on the behavior of the cosmic flow from
decoupling era up to present date in order to answer whether the observed bulk flow is due exclusively to tidal forces.

\section{Analysis of analytic solutions}

In this section, we investigate some analytic solutions of eq.\,(\ref{qvelocity},\ref{t},\ref{Bstar00}) that are obtained thanks to particular
properties of distortion matrix ${\bf B}$. The parameters $\lambda_{\circ}$, $\Omega_{\circ}$ and $H_{\circ}$ given in eq.\,(\ref{Parameters}) correspond
to cosmological Friedmann-Lema\^{\i}tre (FL) world model parameters. Moreover, the constraint $\beta_{2}=0$ in eq.\,(\ref{dt},\ref{Pr}) provides us with
the FL chronology, where $\kappa_{\circ}$ given in eq.\,(\ref{k}) represents the curvature parameter in the FL model ({\em i.e.\/} the dimensionless scalar
curvature
$\Omega_{k}$ of the comoving space, see~\cite{TriayEtAl96}), while the flatness of (simultaneous events) Newton space. It must be noted that the particle
position $\vec{q}$ as defined in eq.\,(\ref{q}) does not identify to the usual FL comoving coordinate because the (generalized) expansion factor $a$
depends on the anisotropy unless $\beta_{2}=0$. 

\subsection{Evolution of functions $\beta_{n=2,3}$}
Because ${\bf B}$ is a traceless matrix, its characteristic polynomial reads
\begin{equation}\label{Polynome}
Q(s)={\rm det}\left(s{\rm 1\mskip-4mu l}-{\bf B}\right)=s^{3}-\frac{1}{2}\beta_{2}s-\frac{1}{3}\beta_{3}
\end{equation}
according to Leverrier-Souriau's algorithm \cite{Souriau92}. With Cayley-Hamilton's theorem ({\it i.e.} $Q({\bf B})=0$) and eq.\,(\ref{Bstar00}) we obtain
the following differential equations system
\begin{eqnarray}
\frac{{\rm d}}{{\rm d}\tau}\beta_{2}&=& -2\beta_{3} \label{TrB2}\\
\frac{{\rm d}}{{\rm d}\tau}\beta_{3}&=& -\frac{1}{2}\beta_{2}^{2} \label{TrB3}
\end{eqnarray}
and we note that the discriminant of third order polynomial $Q$, it is proportional to
\begin{equation}\label{c}
\alpha=3\beta_{3}^{2}-\frac{1}{2}\beta_{2}^{3}
\end{equation}
is a \underline{constant of motion} (i.e., ${\rm d}\alpha/{\rm d}\tau=0$). The integration of eq.\,(\ref{TrB2},\ref{TrB3}) is performed by defining
$\beta_{2}$ by a quadrature
\begin{equation}\label{TraceEvol}
\tau=\tau_{\circ}+\epsilon\frac{\sqrt{6}}{2}\int_{\beta_{2}(\tau_{\circ})}^{\beta_{2}(\tau)}\frac{{\rm d}x}{\sqrt{2\alpha+x^{3}}},\qquad \epsilon=\pm 1
\end{equation}
and hence $\beta_{3}$ from eq.\,(\ref{TrB2}); in addition of the singular solution
\begin{equation}\label{SingularSol}
\beta_{2}=\beta_{3}=0,\qquad (i.e.,\quad {\bf B}^{3}=0)
\end{equation}
defined equivalently either by $\beta_{2}=0$ or $\beta_{3}=0$, according to eq.\,(\ref{TrB2},\ref{TrB3}).

The related dynamics depends on roots $\eta_{i=1,2,3}$ of characteristic polynomial $Q$ given in eq.\,(\ref{Polynome}), {\it i.e.\/} the eigenvalues of
distortion matrix ${\bf B}$. Their real values identify to dilatation rates at time $\tau$ toward the corresponding (time dependent) eigenvectors (not
necessarily orthogonal). Their sum is null ($\beta_{1}=0$) and their product ($\beta_{3}=3{\rm det}{\bf B}$) is either decreasing
with time or is null, according to eq.\,(\ref{TrB3}). The sign of $\alpha$ given in eq.\,(\ref{c}) is used to classify the solutions as follows~:
\begin{itemize}
\item if $\alpha=0$ then $Q$ has a real double root $\eta_{1}=\eta_{2}$ and a simple one $\eta_{3}$. The related instantaneous kinematic shows a
planar-axial symmetry (either a contraction within a plane with an expansion toward a transverse direction or vice versa), see
sec.~\ref{PlanarAxialKinematic}. If $\eta_{1}=\eta_{3}$ then both vanish and the related solution identifies to the singular one defined
in eq.\,(\ref{SingularSol}), see sec.\ref{PlanarKinematic};
\item if $\alpha>0$ then $Q$ has a single real root $\eta_{1}$; 
\item if $\alpha<0$ then $Q$ has three distinct real roots $\eta_{i=1,2,3}$. Their order is conserved during the evolution (since a coincidence of
eigenvalues makes $\alpha=0$), the largest one must be positive while the smallest one must be negative (because $\beta_{1}=0$).
\end{itemize}

\subsection{Planar kinematics}\label{PlanarKinematic}
The singular solution ${\bf B}^{3}=0$ shows a FL chronology and the distortion matrix
\begin{equation}\label{distortionFree}
{\bf B}=-{\bf B}_{\circ}^{2}\tau+{\bf B}_{\circ},\qquad {\bf B}_{\circ}^{3}={\bf 0}
\end{equation}
is solely defined by its initial value ${\bf B}_{\circ}$, according to eq.\,(\ref{Bstar00}). It is neither
symmetric nor asymmetric (otherwise it vanishes), see eq.\,(\ref{amplitude}). Hence, eq.\,(\ref{qvelocity}) transforms
\begin{equation}\label{distortionFreeQ}
\frac{{\rm d}\vec{q}}{{\rm d}\tau}=\left(-{\bf B}_{\circ}^{2}\tau+{\bf B}_{\circ}\right)\vec{q}
\end{equation}
which accounts for eternal motions
\begin{equation}\label{distortionFreeQtriv}
\vec{q}=\exp{\left(-{\bf B}_{\circ}^{2}\frac{\tau^{2}}{2}+{\bf B}_{\circ}\tau\right)}\vec{q}_{\circ}
=\left(1+{\bf B}_{\circ}\tau\right)\vec{q}_{\circ}
\end{equation}
The trajectory of a particle located at initial position $\vec{q}_{\circ}$ identifies to a straight line toward the direction ${\bf
B}_{\circ}\vec{q}_{\circ}$. The analysis of ${\bf B}_{\circ}$ range ({\it i.e.\/}, its image) provides us with characteristics of trajectories
flow. The nilpotent property of ${\bf B}_{\circ}$ shows that its kernel is not empty ${\rm Ker}({\bf B}_{\circ})\neq\emptyset$. Its dimension ${\rm
dim}\left({\rm Ker}({\bf B}_{\circ})\right)=m$ characterizes the kinematics, which is either planar ($m=1$) or directional ($m=2$), {\it i.e.\/} a
bulk flow. Conversely, if the kernel of distorsion matrix ${\bf B}$ is not empty then $\beta_{3}=3\,{\rm det}\left({\bf B}\right)=0$, and thus
$\beta_{2}=0$, see to eq.\,(\ref{TrB2},\ref{TrB3}). Therefore, all planar kinematics can be described by such a model.

\subsection{Planar-Axial kinematics}\label{PlanarAxialKinematic}
If $\beta_{2}\neq0$ then the chronology differentiates from FL one. Let us focus on the $\alpha=0$ with two distinct eigenvalues
$\eta_{1}\neq \eta_{3}$ class of solutions. With eq.\,(\ref{TraceEvol}), eq.\,(\ref{TrB2},\ref{TrB3}) integrate
\begin{equation}\label{TraceBn}
\beta_{n}=\frac{6}{\left(\tau-\tau_{\star}\right)^{n}},\quad (n=2,3), \qquad \tau_{\star}=\tau_{\circ}+\sqrt{6}\beta_{2}^{-1/2}(\tau_{\circ})
\end{equation}
which shows a singularity at date $\tau=\tau_{\star}>0$ that splits the motion in two regimes $\tau<\tau_{\star}$ and $\tau>\tau_{\star}$. The complete
investigation of this singularity problem demands to solve an integro-differential equation, see eq.\,(\ref{dt},\ref{Pr}). The roots of $Q$
read
\begin{equation}\label{C0ValP}
\eta_{1}=\frac{1}{\left(\tau_{\star}-\tau\right)},\qquad
\eta_{3}=-2\eta_{1}
\end{equation}
where $\eta_{1}$ stands for the double root. Among others, two class of solutions are defined by mean of a constant (time independent) matrix ${\bf P}$,
the projector associated to $\eta_{1}$, see~\cite{Souriau92},
\begin{equation}\label{P1}
{\bf P}^{2}={\bf P},\quad
{\rm tr}{\bf P}=2
\end{equation}
They describe distinct kinematics depending on whether matrix ${\bf B}$ is diagonalizable.

\subsubsection{(Irrotational) motions}
If ${\bf B}$ is diagonalizable then
\begin{equation}\label{PancakeB}
{\bf B}=\eta_{1}\left(3{\bf P}-2{\rm 1\mskip-4mu l}\right)
\end{equation}
From eq.\,(\ref{qvelocity},\ref{PancakeB}), one has
\begin{equation}\label{Caust_q_vel}
\frac{{\rm d}\vec{q}}{{\rm d}\tau}=\eta_{1}\left(3{\bf P}-2{\rm 1\mskip-4mu l}\right)\vec{q}\\
\end{equation}
and the solution reads
\begin{equation}\label{irrotational}
\vec{q}= -\eta_{1}{\bf P}\vec{\xi}+\frac{1}{\eta_{1}^{2}}\left({\rm 1\mskip-4mu l}-{\bf P}\right)\vec{\xi}
\end{equation}
where $\vec{\xi}$ is constant. If the eigenvectors are orthogonal then the kinematics accounts for irrotational motions. 

\subsubsection{Rotational motions}
If ${\bf B}$ is not diagonalizable then
\begin{eqnarray}
{\bf B}&=&\eta_{1}\left(3{\bf P}-2{\rm 1\mskip-4mu l}\right)+\frac{1}{\eta_{1}^{2}}{\bf N}\label{PancakeB1}\\
\frac{{\rm d}\vec{q}}{{\rm d}\tau}&=&\left(\eta_{1}\left(3{\bf P}-2{\rm 1\mskip-4mu l}\right)+\frac{1}{\eta_{1}^{2}}{\bf N}\right)\vec{q}\label{PancakeB2}
\end{eqnarray}
where ${\bf N}$ is a constant nilpotent matrix, which accounts for rotational motions on the eigenplane of ${\bf P}$. The
solution reads
\begin{equation}\label{rotational}
\vec{q}= -\eta_{1}{\bf P}\vec{\xi}+\frac{1}{\eta_{1}^{2}}\left({\rm 1\mskip-4mu l}-{\bf P}\right)\vec{\xi}+\frac{1}{\eta_{1}^{2}}{\bf N}\vec{\xi}
\end{equation}
where $\vec{\xi}$ is constant.

\section{Application to flat large scale structures (${\bf B}^{3}=0$)}
The ${\bf B}^{3}=0$ class of solutions has interesting properties with regard to the stability of large scale structures that show a flat spatial
distribution. To answer the question of whether observations define unambiguously the kinematics, the distortion matrix ${\bf B}$ is decomposed as follows
 \begin{equation}\label{Bdecomp}
{\bf B}={\bf S}+{\bf j}(\vec{\omega}),\qquad {\rm tr}{\bf S}=0
\end{equation}
where ${\bf S}$ and ${\bf j}(\vec{\omega})$ stand for its symmetric and its asymmetric\footnote{The operator ${\bf j}$
stands for the vector product, $\vec{u}\times\vec{\omega}={\bf j}(\vec{u})(\vec{\omega})$.} component, and
\begin{equation}\label{tourbillon}
\vec{\omega}=\frac{a^{2}}{H_{\circ}}\vec{\sigma},\qquad \vec{\sigma}=\frac{1}{2}\vecp{\rm rot}\vec{v}
\end{equation}
accounts for the motion rotation, $\vec{\sigma}$ being the swirl vector.  Hence, eq.\,(\ref{distortionFree})
gives
\begin{equation}\label{tourbillonV}
{\bf B}\vec{\omega}={\bf S}\vec{\omega}
\end{equation}
The evolution of the anisotropy with time is defined by
\begin{eqnarray}
{\bf S}&=&-\left({\bf S}_{\circ}^{2}+ {\bf j}(\vec{\omega_{\circ}}){\bf j}(\vec{\omega_{\circ}})\right)\tau+{\bf S}_{\circ}\label{CinPlane1}\\
{\bf j}(\vec{\omega})&=&-\left({\bf S}_{\circ}{\bf j}(\vec{\omega_{\circ}})+{\bf j}(\vec{\omega_{\circ}}){\bf S}_{\circ}\right)\tau+{\bf
j}(\vec{\omega_{\circ}})\label{CinPlane2}
\end{eqnarray}
which couples the symmetric and the antisymmetric parts of the distortion matrix. The swirl magnitude reads
\begin{equation}\label{tourbillonM}
\omega=\sqrt{\langle\vec{\omega},\vec{\omega}\rangle}=\sqrt{\frac{1}{2}{\rm tr}{\bf S}^{2}}
\end{equation}
according to eq.\,(\ref{tourbillon}), since $\beta_{2}=0$. Its orientation cannot be determined from the data because the above equations describe two
distinct kinematics corresponding to $\pm\vec{\omega_{\circ}}$ that cannot be disentangle. According to eq.\,(\ref{Bdecomp}), if (and only if) the
rotation $\omega=0$ then the distortion vanishes ${\bf S}={\bf 0}$ since ${\bf B}$ is either a symmetric or antisymmetric. In other words, a planar
distortion has necessarily to account for a rotation.

\subsection{Constant distortion}
Among above solutions which show planar kinematics, let us investigate the (simplest) one defined by ${\bf
B}^{2}_{\circ}={\bf 0}$. In such a case, $\vec{k}_{\circ}\propto\vec{\omega}$ and ${\bf S}\vec{\omega}=\vec{0}$.  According
to eq.\,(\ref{CinPlane1},\ref{CinPlane2},\ref{tourbillonM}), linear calculus shows that the distortion is constant
\begin{equation}\label{tourbillonF}
{\bf S}={\bf S}_{\circ},\qquad \omega=\omega_{\circ}
\end{equation}
Such a distortion in the Hubble flows produces a rotating planar velocities field with magnitude $\propto H_{\circ}a^{-2}$. In the present case, the
model parameters can be easily evaluated from data. The observed cosmic velocity fields are partially determined by their radial component
\begin{equation}\label{VitRad}
v_{r}=\langle\vec{v},\frac{\vec{r}}{r}\rangle=cz,\qquad \vec{r}=\vec{r}(m)=r\vec{u}=a\vec{q},\qquad r=ct
\end{equation}
where $m$, $z$, $\vec{u}$, $t$ stand respectively for the apparent magnitude, the redshift, the line of sight, the photon emission date of
the galaxy and $c$ the speed of the light. According to eq.\,(\ref{velocity},\ref{H},\ref{Bdecomp}), the radial velocity of a galaxy
located at position $\vec{r}$ is given by
\begin{equation}\label{Vr}
v_{r}=\left(H +\tilde{H}_{\vec{u}}\right)r,\qquad \tilde{H}_{\vec{u}}=\frac{H_{\circ}}{a^{2}}\vec{u}\cdot{\bf
S}\vec{u}
\end{equation}
Because ${\rm tr}{\bf S}=0$, it is clear that the sum of three radial velocities $v_{r}$ corresponding to galaxies located in the sky toward
orthogonal directions and at same distance $r$ provides us with the quantity $H$. Hence, simple algebra shows that the sample average of radial
velocities within a sphere a radius $r$ is equal to
\begin{equation}\label{Hsample}
\langle v_{r}\rangle=H\langle \vec{r}\rangle
\end{equation}
Therefore, for motions described by eq.\,(\ref{HubbleLaw}), the statistics given in eq.\,(\ref{Hsample}) provides us with a genuine interpretation
of Hubble parameter $H=H(t)$. Hence, according to eq.\,(\ref{H},\ref{dt},\ref{Pr}), one obtains the (generalized) expansion factor
\begin{equation}\label{at}
a(t)=\exp{\int_{t_{\circ}}^{t}H(t){\rm d}t}
\end{equation}
Hence, the cosmological parameters can be estimated by fitting the data to the function 
\begin{equation}\label{psi}
\psi(t)_{\lambda_{\circ},\kappa_{\circ},\Omega_{\circ}}=\sqrt{P(a)}=a^{2}H/H_{\circ},\quad \lambda_{\circ}+\kappa_{\circ}+\Omega_{\circ}=1
\end{equation}
The component of matrix ${\bf S}$ can be estimated by substituting $H$ in eq.\,(\ref{Vr}), and $\omega$ is obtained from eq.\,(\ref{tourbillonF}).

It is clear that the above model is derived in the Galilean reference frame, where the Euler-Poisson equations system can be applied. Hence, a non
vanishing velocity of the observer with respect to this frame an produces a bipolar harmonic signal in the $\tilde{H}_{\vec{u}}$ distribution
of data in the sky, which can be (identified and then) subtracted. 

\subsection{Discussion}
The dynamics of a homogenous medium and anisotropic moving under Newton gravity was already studied by describing the evolution of an
ellipsoid \cite{ZeldovichNovikov83}. The current approach enables us to identify characteristics of the dynamics of the deformation from isotropic Hubble
law in a more systematic way by mean of the distorsion matrix.

At first glance, if the planar anisotropic of the space distribution of galaxies within the Local Super Cluster (LSC) is stable then the above
solution can be used for understanding its cosmic velocity fields, $\vec{k}_{\circ}$ being orthogonal to LSC plane. It is well known however that
the distribution of galaxies is not so homogenous as that, whereas this model describes motions of an homogenous distribution of gravitational
sources. However, such an approximation level is similar to the one which provides us with the observed Hubble law, that is included in this model
(${\bf S}=0$).

\section{Conclusion}\label{Conclusion}
The present anisotropic solution of Euler Poisson equations system generalizes the Hubble law and provides us with a better understanding of cosmic
velocity fields within large scale structures as long as Newton approximation is valid. As a result, this generalization of Hubble
motion implies necessarily an homogenous distribution of gravitational sources, as similarly to Hubble law. Because the chronology identifies to FL
chronology for a vanishing distortion, this model interprets as a Newton approximation of anisotropic cosmological solutions. The motions are characterized
by means of a constant of motion
$\alpha$. Among them, particular solutions can be easily derived for $\alpha=0$. They describe all planar distortions, in addition of two classes of
planar-axial distortions with or without rotation. Among these solutions, the one which ensures a planar kinematics is of particular interest because it
describes constant (eternal) and rotational distortions. This solution can be fully determined from observational data except for the orientation of the
rotation. The sensible result is that the velocity field is not potential. It is interesting to note that this model accounts for motions which might be interpreted as due to tidal forces whereas the density is homogeneous. It is
an alternative to models which assume the presence of gravitational structures similar to Great Attractor as origin of a bulk flow.


\section*{References}



\end{document}